\documentclass[conference]{IEEEtran}

\usepackage{cite}
\usepackage{t1enc}
\usepackage{url}
\usepackage[cmex10]{amsmath}
\usepackage{amsthm}

\usepackage[pdftex]{graphicx}
\graphicspath{{diagrams/}{}}
\DeclareGraphicsExtensions{.pdf,.png,.eps,}

\newenvironment{definition}[1][Definition]{\begin{trivlist}
\item[\hskip \labelsep {\bfseries #1}]}{\end{trivlist}}
\newenvironment{example}[1][Example]{\begin{trivlist}
\item[\hskip \labelsep {\bfseries #1}]}{\end{trivlist}}

\begin{document}

\title{Software is a directed multigraph\\(and so is software process)}

\author{\IEEEauthorblockN{Robert D\k{a}browski, Krzysztof Stencel, Grzegorz Timoszuk}
\IEEEauthorblockA{Institute of Informatics\\
Warsaw University\\
Banacha 2, 02-097 Warsaw, Poland\\
Email: \{robert.dabrowski, krzysztof.stencel, grzegorz.timoszuk\}@mimuw.edu.pl}
}

\maketitle

\begin{abstract}
For a software system, its architecture is typically defined as the fundamental organization of the system incorporated by its components, their relationships to one another and their environment, and the principles governing their design. If contributed to by the artifacts coresponding to engineering processes that govern the system's evolution, the definition gets natually extended into the architecture of \emph{software} and \emph{software process}.

Obviously, as long as there were no software systems, managing their architecture was no problem at all; when there were only small systems, managing their architecture became a mild problem; and now we have gigantic software systems, and managing their architecture has become an equally gigantic problem (to paraphrase Edsger Dijkstra).

In this paper we propose a simple, yet we believe effective, model for organizing architecture of software systems.

First of all we postulate that only a hollistic approach that supports continuous integration and verification for all \emph{software} and \emph{software process} architectural artifacts is the one worth taking.
Next we indicate a graph-based model that not only allows collecting and maintaining the architectural knowledge in respect to both \emph{software} and \emph{software process}, but allows to conveniently create various quantitive metric to asses their respective quality or maturity.
Such model is actually independent of the development methodologies that are currently in-use, that is it could well be applied for projects managed in an adaptive, as well as in a formal approach.
Eventually we argue that the model could actually be implemented by already existing tools, in particular graph databases are a convenient implementation of architectural repository.
\end{abstract}

\IEEEpeerreviewmaketitle

\section{Introduction}

Nowadays software systems are being developed by teams changing over time, working under time pressure in peek activity periods, working over incomplete documentation and changing requirements, integrating unfamiliar source-code in multiple development technologies, programming languages, coding standards, productively delivering only partially completed releases in iterative development cycles.

When at some point development issues arise (bugs, changes, extensions), they frequently lead to refactoring of the software and the software process. Even if promptly addressed, due to volatile development team leading to insufficient flow of information and inability to properly manage knowledge on software and software process architecture, the problems recurr and preserve in consecutive releases.

Unsurprisingly such challenges had already been identified and software engineering has been focusing on their resolution.

In particular a number of software development methodologies (ie. structured, iterative, adaptive), design models (ie. erd, dfd, std), development languages (ie. functional, object-oriented, domain-specific) and production management tools (ie. issue trackers, build and configuration managers, source-code analysers) have emerged. Although they address important areas, it still remains a challenge for a single project to integrate those methodologies, standards, languages, metric, tools into a consistent platform providing a complete and systematic environment that includes all software and software process artifacts, identifies their dependencies, allows for a systematic build of deliverables, and preveils over development team changes - that is encapsulates and preserves all information important to ensure quality of the system being developed and the predictability of its development process.

For software practitioners this current lack of integration of architectural knowledge is a historical condition: while software was limited to a small number of files delivered in one programming language and built into a single executable, it was possible to browse the artifacts in a \emph{list} mode (file by file; or procedure by procedure). Next, as software projects evolved to become more complex and sophisticated, the idea of a software project organized according to a \emph{tree} (folders, subfolders and files; or classes, subclasses and methods) emerged to allow browsing artifacts in a hierarchical approach.

This is no longer enough. We believe that although software engineering is pursuing the correct direction, the research will lack proper momentum without a new sound model to support integration of current trends, technologies, languages. A new vision for software and software process architecture is required and this paper aims to introduce one and trigger a discussion.

The vision can be summarized as follows: \emph{all} \emph{software} and \emph{software process} artifacts being created during a software project are organized as vertices of a \emph{graph} (being step next after the \emph{list} and \emph{tree}) connected by multiple edges that represent multiple kinds of dependencies among those artifacts. The key aspects of software production, being: the assurance of the \emph{quality} of software; the \emph{predictability} of software development process; the \emph{automation} level of development tools; are enforced by multiple \emph{metric} that \emph{integrate} software and software process artifacts, and are easily expressed and calculated in graph terms.

Presently we are only at the beginning of a road to such vision and the theoretical foundations, the range of supporting tools, and the extension of its systematic evaluation require a significant amount of further research.

The rest of the paper is organized as follows. In Section 2 we analyze the background that motivated our approach. In Section 3 we provide a definition of the graph-based model for architectural knowledge management. In Section 4 we indicate that the model could be implemented with existing tools and define challenges for further research.

\section{Motivation}

In this paper we postulate that in the coming years software engineering would be imminent to focus on providing means (both theoretical foundations and supporting tools) for the following vision of software development process: \emph{all} artifacts created in a software project are organized according to a consistent graph-based model in which the dependencies between those artifacts (and therefore the respective flow of information) are ensured by providing each significant artifact type with a clear specification that the artifacts are restricted to comply to (ie. artifact-specific language), accompanied by a set of tools organized in tool-chains that provide automated transition between those artifacts.
Actually such approach is not an entirely novel one, but should rather be perceived as being contributed to by several trends and approaches already existing in software engineering.

The starting point for this perception is the strive for quantitative assessment of software quality and software process predictability. Typically this gets achieved with different metric. While the mathematical nature of metric requires clarification, frequently there exist many contradicting definitions of the same metric (ie. depending on the implementation language). It has been suggested by Mens and Lanza \cite{ENTCSvol72no2y2002} that metric should be expressed and defined using a language-independent metamodel based on graphs. Such graph-based approach allows for an unambiguous definition of generic object-oriented and higher-order metric.

Also Gossens, Belli, Beydeda and Dal Cin \cite{1112121} considered view graphs for representation of source code convenient for program analysis and testing at different levels of abstraction (ie. white-box analysis and testing at the low level of abstraction; black-box analysis and testing at the high level of abstraction). Graph-based approach thus integrates the different analysis and testing techniques.

Nowadays it has became frequent that models of software often describe systems by a number of (partially) orthogonal views (ie. a state machine, a class diagram, a scenario might specify different aspects of the given system being built). Such abstract, multi-view models are the starting point for transformations into platform-specific models and finally the code. However, during these transformations it is usually not possible to keep such a neat separation into different views: the specification language of the target models might not support all such views. The target model, however, still needs to preserve the behavior of the abstract, multi-view model. Therefore, model transformations have to be capable of moving aspects of the behavior across views.
Derrick and Wehrheim \cite{DBLP:journals/scp/DerrickW10} studied aspects of model transformations migrations from state-based views (ie. class specifications with data and methods) to protocol-based views (ie. process specifications on orderings of methods) and vice versa. They suggested specification languages for these two views be equipped with a joint, formal semantics which enables a proof of behavior preservation and consequently derives conditions for the transformations to be behavior-preserving. 

Also Fleurey, Baudry, France and Ghoshit \cite{1423617} have observed that it is necessary to have the ability to automatically compose models to build a global view of the system. The graph-based approach allows for a generic framework for model composition that is independent from a modeling language.

The use of components in development of complex software systems surely has various benefits, however their testing is still one of the top issues in software engineering. In particular both the developer of a component and the developer of a system, while using components, often face the problem that information vital for certain development tasks is not available.
Such lack of information has various consequences to both. One of the important consequences is that it might not only obligate the developer of a system to test the components used, it might also complicate these tests. Beydeda and Gruhn \cite{10.1007/978-3-540-69073-3_2} have focused on component testing approaches that explicitly respect this lack of information during development.

As K{\"u}hne, Selic, Gervais and Terrier \cite{DBLP:conf/ecmdafa/2010} have noticed, the automated support for the transition from use cases to activity diagrams would provide significant, practical help. Additionally, traceability could be established through automated transformation, which could then be used to relate requirements to design decisions and test cases. They proposed an approach to automatically generate activity diagrams from use cases while establishing traceability links.

Osterweil \cite{Osterweil:1987:SPS:41765.41766} has long been startled to realize that software systems are large, complex and intangible objects developed without a suitably visible, detailed and formal descriptions of how to proceed. He suggested that not only the software, but also software process should be included in software project as programs with explicitly stated descriptions.
According to Osterweil, the manager of a project should communicate to developers, customers and other managers through a software process program, indicating just what steps are to be taken in order to achieve product development or evolution goals. Developers, in particular, can benefit from software process programs in that reading them should indicate the way in which work is to be coordinated and the way in which each individual's contribution is to fit with others' contributions. In this sense software process program is another of artifacts in the graph we propose in this paper.

\section{Model}

In the remaining part of the chapter we introduce a model based on \emph{directed multigraphs} to represent software and software process architecture.

\begin{definition}
Let $\cal S$ be a software-intensive system. Let $\cal A$ denote the set of all types of artifacts that get created during construction of $S$, let $\cal T$ denote the set of all types of traces (or dependencies) among those artifacts. In the remaining part we assume $\cal A, \cal T$ to be given and denote $\cal S = \cal S (\cal A, \cal T)$.
\end{definition}

The set $\cal A$ is a dictionary of attributes that annotate artifacts created during development of $\cal S$. For the simplicity of reasoning we assume $\cal A$ to be predefined in the remaining part of the paper. It should be clearly stated that it remains a challenge to derive a representative and consistent classifcation (a superset) of such atrributes, since during development of an actual software only a subset of $\cal A$ is typically used. On the other hand, the scope of artifact types used in a given project may be an incentive for new metric to assess its production process maturity.

\begin{example}
Typically $\cal A$ may contain some of the following values:
\begin{itemize}
	\item class
	\item coding standard
	\item field
	\item grammar
	\item interface
	\item library
	\item method
	\item module
	\item requirement
	\item test suite
	\item use case
	\item unit test
	\item $\dots$
\end{itemize}
\end{example}

Analogically, the set $\cal T$ should be perceived as the dictionary of labels to describe relationships traced among the artifacts. Again, in the remaining part of the chapter we assume it to be predefined, although the set of actual traces may be software-specific and derivation of a common superset remains a challenge.

\begin{example}
Typically $\cal T$ may contain some of the following values:
\begin{itemize}
	\item apply to
	\item call
	\item contain
	\item define
	\item depend on
	\item generate
	\item implement
	\item limit
	\item require
	\item return
	\item use
	\item verify
	\item $\dots$
\end{itemize}
\end{example}

\begin{definition}
The \emph{software graph} $\cal G$ is an ordered tuple $\cal G (\cal S) = (\cal V, \cal L, \cal E)$, where $\cal V$ is the set of vertices that represent the software or software process artifacts, $\cal L \subseteq \cal V \times \cal A$ is the annotation of vertices with their attributes, ${\cal E} \subseteq {\cal V} \times {\cal T} \times {\cal V}$ is the set of directed edges that trace dependencies between artifacts.
\end{definition}

$\cal G$ is a multigraph, that is there can be more than one edge in $\cal E$ from one vertex in $\cal V$ to another vertex in $\cal V$.
$\cal G$ is a directed graph, that is forward and backward relations traced among artifacts are distinguished. 

\begin{example}
Typically $\cal E$ may contain some of the following values:
\begin{itemize}
	\item a class \emph{calls} a class
	\item a class \emph{contains} a field
	\item a class \emph{contains} a method
	\item a class \emph{implements} an interface
	\item a coding standard \emph{limits} a module
	\item a grammar \emph{generates} a class
	\item a method \emph{calls} a method
	\item a module \emph{depends on} a module
	\item a requirement \emph{defines} a module
	\item a unit test \emph{verifies} a method
	\item $\dots$
\end{itemize}
\end{example}

The graph model defines a foundation that integrates all software and software process artifacts created during software production. However, the model may grow large for a non-trivial software. For the architectural information to remain accessible, the model must provide means for a human-convenient representation.

For this purpose we define a \emph{graph view} that subsets the graph to a given scope of artifacts and their traces.

\begin{definition}
For a given software graph $\cal G = ({\cal V}, {\cal L}, {\cal E})$ and subsets of its artifact types $\cal A' \subseteq \cal A$ and traces types $\cal T' \subseteq \cal T$, its \emph{view} is a subgraph ${\cal G}|_{\cal A',\cal T'} = ( \cal V', \cal L', \cal E' )$, where
${\cal V'} = \{ v \in {\cal V} | \exists_{a \in \cal A'} (v,a) \in {\cal L} \}$,
${\cal L'} = \cal V' \times \cal A'$ and
${\cal E'} = {\cal E} \cap ({\cal V'} \times {\cal L'} \times {\cal V'})$.
\end{definition}

In the following examples we include sample figures extracted from an actual project mapped onto the model being postulated in this paper. Because of the limited scope of the paper, we retain the figures for demonstrative purposes without any detailed explanations.

\begin{example}
The \emph{class view} of a given software graph $\cal G$ is its subgraph ${\cal G}|_{A',T'}$ where
$\cal A'=\{$ class, interface, method, field $\}$,
$\cal T'=\{$ contain, implement, return $\}$.\\

\centering
\includegraphics[width=2.7in]{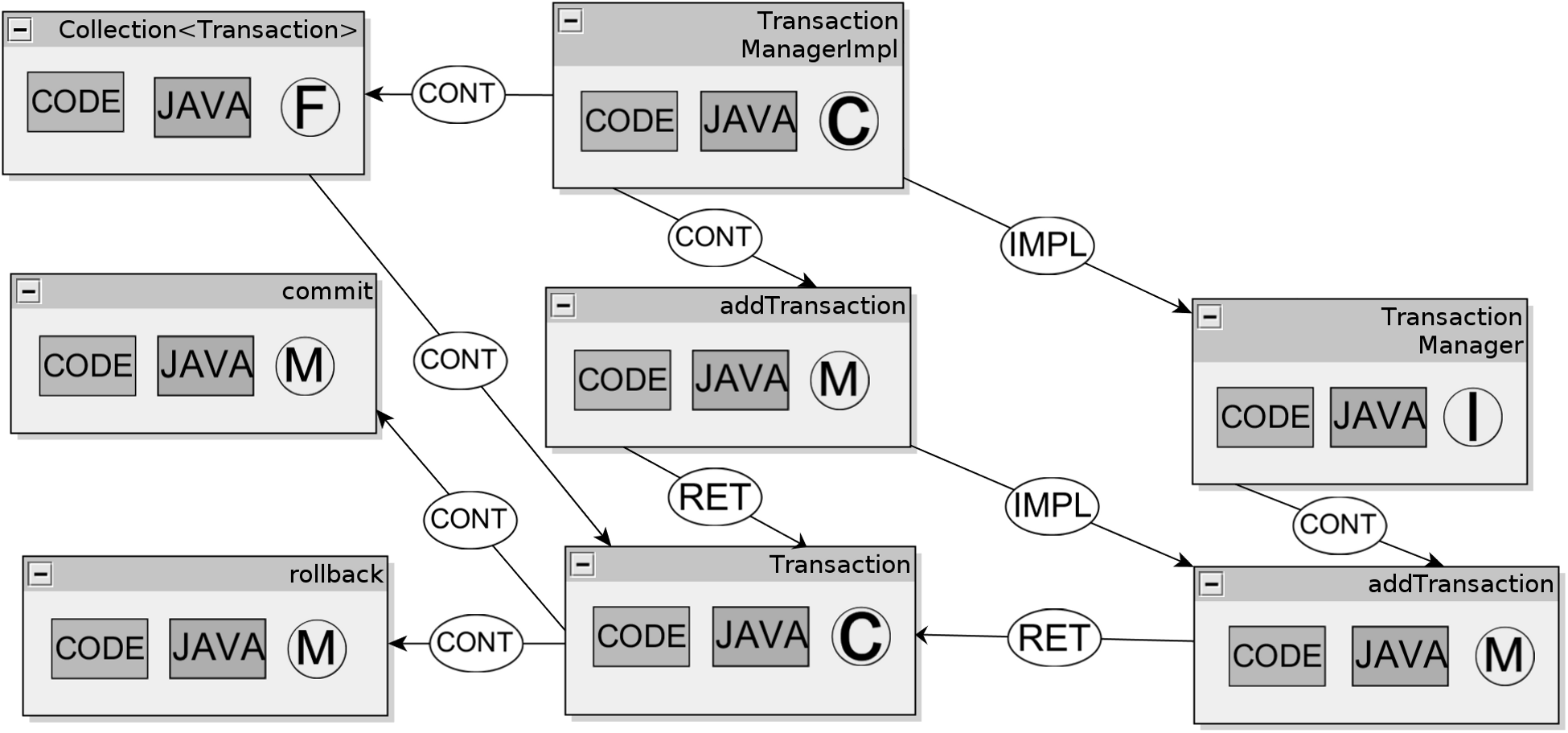}\\
{\bfseries Fig} A class view (sample)
\end{example}

Sometimes a more abstract perspective on the software and software process artifacts is required. In particular it may be beneficiary for a software architect or a software developer to quit distinguishing certain differences in artifact or trace types (ie. vertices can be simply nested in one another, like in case of a class that contains a field or a method). For this purpose we introduce the notion of a \emph{graph map}. Please note that the graph map combined with graph view gets natually usefull when drawing a human-convenient graphical (ie. two or three-dimensional) representation of a given software graph.

\begin{definition}
(intentional) For a given software graph $\cal G = ({\cal V}, {\cal L}, {\cal E})$ and a transformation on its edges $t: {\cal T} \times {\cal T} \mapsto {\cal T}$,
its \emph{map} is a subgraph ${\cal G}^{t} = \{ \cal V, \cal L, {\cal E}'\}$, where 
$\cal E'$ is the set of new edges resulting from a transitive closure of mapping $t$ calculated on the neighbouring edges of vertices in $\cal G$.
\end{definition}

\begin{example}
The classical \emph{class diagram} of a given software graph $\cal G$ is its subgraph ${\cal G}^{t}|_{A,T}$ where
$t: \{$ contain, return $\} \mapsto \{$ depend $\}$,
$\cal A'=\{$ class $\}$,
$\cal T'=\{$ depend $\}$.\\

\centering
\includegraphics[width=2.7in]{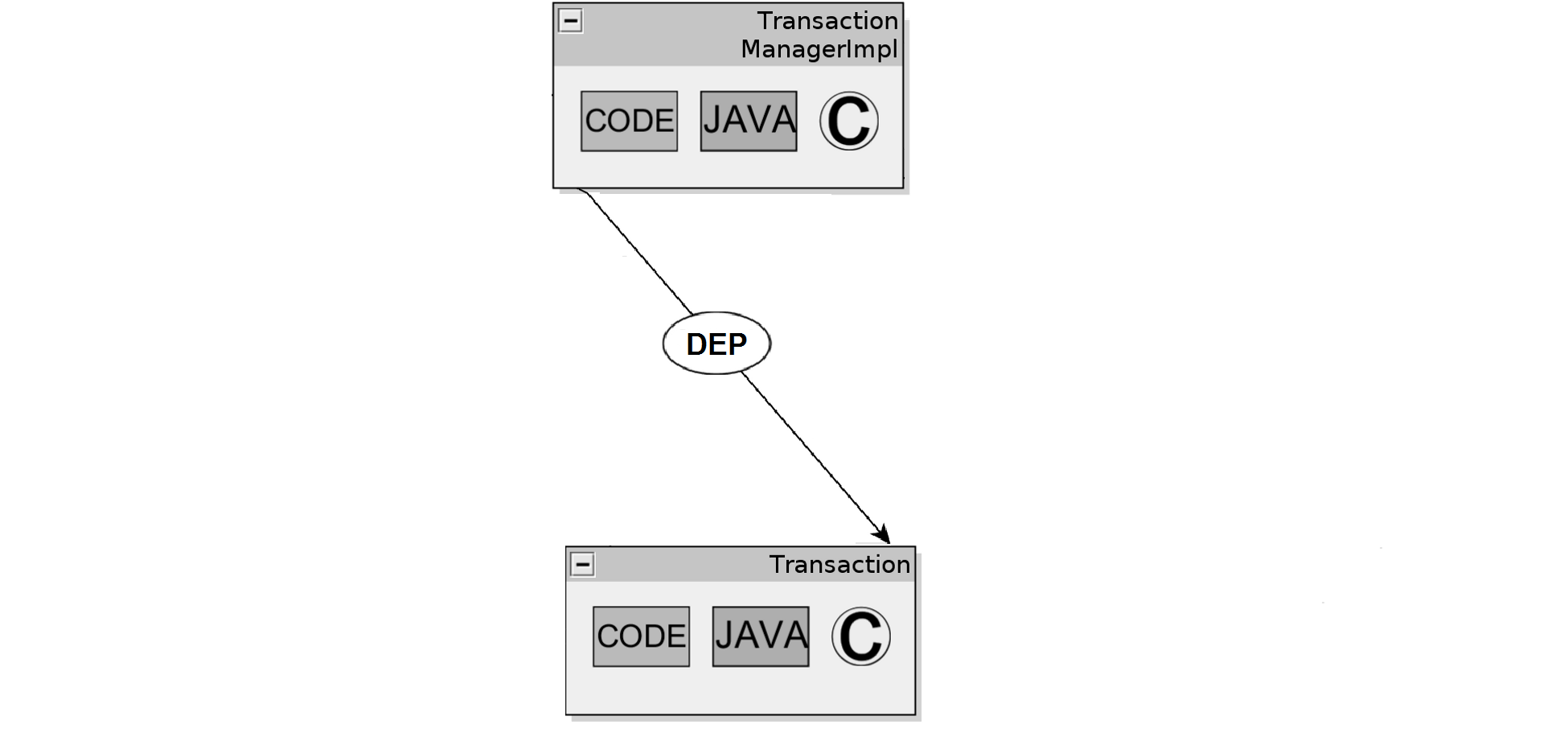}\\
{\bfseries Fig} A class diagram (sample)
\end{example}

Eventually it is important to note that such graph-based approach allows not only using existing metric that can be efficiently calculated using graph algorithms \cite{Hopcroft:1973:AEA:362248.362272}, but also allows designing new metric. Particularly interesting would be metric that integrate both software and software process artifacts.

\begin{definition}
(intentional) For a given software graph $\cal G (\cal A, \cal T) = ({\cal V}, {\cal L}, {\cal E})$, its \emph{metric} is any function $m: $\cal A, \cal T, \cal V, \cal L, \cal E$ \mapsto \cal R$ ($\cal R$ being real numbers) which can be efficiently calculated by an algorithm that takes as an input $\cal A, \cal T, \cal V, \cal L, \cal E$.
\end{definition}

\begin{example}
Counting \emph{neighbouring vertices} of selected type allows to easily calculate different \emph{coupling factors}. Graph property of \emph{reachability} induces a number of easily assesed software qualities: in case of \emph{requirements} verices it can be expected that the vertices of all other types are reachable from at least one \emph{requirement}; to assess \emph{test coverage} it can be expected that each \emph{method} is reachable from at least one \emph{unit test}.
\end{example}

\section{Conclusion}

We follow the research on architecture of software \cite{Kruchten:2005:BUE:1130239.1130708} and software process, and we believe that only an approach that avoids separation between software and software process artifacts is the one worth taking \cite{Osterweil:1987:SPS:41765.41766}. Implementation of such approach has already became feasible - starting with a graph-based model and using graph databases \cite{anglesgutierres} as the foundation for artifact representation.

The concept is not an entirely novel one, rather it should be perceived as an attempt to support existing trends with a sound and common foundation. We believe a hollistic approach is required for current research to gain proper momentum, as despite many advanced tools, actual software projects still suffer from a lack of visible, detailed and complete setting to govern their architecture and evolution.

We are also aware that the scope of research required to turn this vision into an actual contribution to software engineering requires further work.
In particular we believe the following research areas to be especially interesting:
defining UML diagrams as reports obtained from the integrated software graph as a combination of its views and maps;
designing an actual implementation of the graph based on graph databases; designing a query language that would operate on the graph model and allow architects and developers to conveniently filter, zoom, drill-down the architectural information;
assesing a representative number of existing projects in an effort to provide a systematic classifications of artifacts types $\cal A$ and traces types $\cal T$; perhaps the artifacts types and traces types should evolve rather to be trees then mere lists;
designing metric (in graph-based terms, so they can be calculated by graph algorithms) to assess software quality and software process maturity; implementing graph algorithms to calculate those metrics;
precise definitions for the model and its components (views, maps), new components enriching the model;
classifying existing software and its process according to the model, in particular calculating metric in order to assess software quality and software process maturity; this would allow to compare the software projects to one another.

\bibliographystyle{./IEEEtran}
\bibliography{./IEEEabrv,./graph,./biblio}

\end{document}